\documentstyle[epsfig,revsymb]{article}

\title{ELECTROMAGNETIC AND GRAVITATIONAL RADIATION OF GRAVIATOMS}

\author
   {Yu.P. Laptev\footnote{yplaptev@rambler.ru} and M.L. Fil'chenkov\footnote{fmichael@mail.ru}\\\\
Institute of Gravitation and Cosmology, \\
Peoples' Friendship University of Russia\\ 6 Miklukho-Maklaya
Street, Moscow 117198, Russia}

\begin{document}

\maketitle

\begin{abstract}
Graviatom existence conditions have been found. The graviatoms
(quantum systems around mini-black-holes) satisfying these
conditions contain the following charged particles: the electron,
muon, tau lepton, wino, pion and kaon. Electric dipole and
quadrupole and gravitational radiations are calculated for the
graviatoms and compared with Hawking's mini-hole radiation.
\end{abstract}


\section{Introduction}

A motion of micro-particles on scales larger than the Compton
length is quantized in curved space-time. Kucha\v{r} has shown
that the latter reduces to Schr\"odinger's equation with the
Newtonian potential in a non-relativistic case \cite{K}. The
behaviour of charged particles in a centrally symmetric
gravitational field was considered by DeWitt who obtained a
so-called self-force, acting on the charge, apart from the
Newtonian gravitational force \cite{D}.

A quantum-mechanical problem of electron motion in the gravitational field
of a mini-hole was considered by Gaina who obtained hydrogen-like
solutions \cite{G}.

The general case of charged particle motion in the Schwarz\-schild field
was considered later taking account of DeWitt's self-force \cite{F1,F2,LF}.
Primordial black holes (or mini-holes) can capture charged particles due to
gravitational interaction. Bound quantum systems maintaining a charged
particle in orbit around a mini-hole were called graviatoms \cite{LF}.

In the present article, we shall derive graviatom existence
conditions and calculate electromagnetic and gravitational
radiations to be compared with Hawking's mini-hole radiation.

\section{Academic problem solution}

Schr\"odinger's equation for the graviatom \cite{F1}
$$
   \frac{1}{r^2}\frac{d}{dr}\left[r^2\left(\frac{dR_{pl}}{dr}\right)\right]
        -\frac{l(l+1)}{r^2}R_{pl}+\frac{2m}{\hbar^2}\left(E-\frac{mc^2r_Qr_g}{4r^2}
        +\frac{mc^2r_g}{2r}\right)R_{pl} = 0 \eqno(1)
$$
describes a radial motion of a particle with the charge $Q$ and
mass $m$ in the effective mini-hole potential, taking into account
DeWitt's self-interaction, as follows
$$
  U_{\rm eff}=-\frac{mc^2r_{r_g}}{2r}+\frac{mc^2r_Qr_g}{4r^2}+\frac{\hbar^2
        l(l+1)}{2mr^2}, \eqno(2)
$$
 where $r_g = 2GM/c^2$ and $M$ are the mini-hole gravitational
radius and mass, respectively, and $r_Q = Q^2/(mc^2)$ is the
classical radius of the charged particle.

The solution to (1) has the form
$$
    R_{\rm pl} = const\cdot \rho^s e^{-\rho/2}F(-p, 2s+2, \rho), \eqno(3)
$$
 where $F(-p, 2s+2, \rho)$ is the confluent hypergeometric
function,
$$
    \rho=\frac{2\sqrt{-2mE}}{\hbar}r,\qquad
            s(s+1)=\frac{2mA}{\hbar^2}+l(l+1),
$$
$$
    A=\frac{mc^2r_Qr_g}{4},\qquad B=\frac{mc^2r_g}{2},\qquad p=n-s-1,
$$

 $p=0,1,2,...,\quad l\le n, \quad n=1,2,3,...\quad n$ and $l$ are
the principal and orbital quantum numbers, respectively.

The energy spectrum of the charged particle captured by the
mini-hole reads
$$
    E=-\frac{2B^2m}{\hbar^2}\frac{1}{\left[2p+1+\sqrt{(2l+1)^2
            +\frac{8mA}{\hbar^2}}\right]^2}.           \eqno(4)
$$

If we take account of only the electromagnetic and gravitational
interactions and neglect the strong one, determining nuclei sizes,
the solutions (3) and (4) for the nuclei with the mass $m=2Zm_p$
and charge $Q=Ze$ can be divided into two cases: the hydrogen-like
one for light nuclei with $r_gr_Q/\lambdabar_c^2\ll 1$, whose
energy spectrum is given by the formula
$$
    E = -\frac{4Z^3m_p^3G^2M^2}{\hbar^2 n^2},\eqno(5)
$$
 where $\lambdabar_c = \hbar/(mc)$ is the Compton wavelength,
$m_p$ is the proton mass and $G$ the gravitational constant, and
that for Kratzer's potential, valid for heavy nuclei with $r_g
r_Q/\lambdabar_c^2 \gg 1$, whose energy spectrum takes the form
$$
    E=-\frac{mc^2r_g}{4r_Q}+\frac{\hbar
    c}{r_Q}\left(p+\frac{1}{2}\right)\sqrt{\frac{r_g}{2r_Q}}
        +\frac{\hbar^2}{2mr_Q^2}\left(l+\frac{1}{2}\right)^2 - \frac{3\hbar^2}{2mr_Q^2}\left(p+\frac{1}{2}\right)^2-
    \frac{3\hbar^3}{m^2cr_Q^2\sqrt{2r_gr_Q}}
    \left(p+\frac{1}{2}\right)\left(l+\frac{1}{2}\right)^2,       \eqno(6)
$$
 where the second term describes oscillations, the third
rotations, the fourth anharmonicity of oscillations and the fifth
oscillation-rotation coupling. Below we shall call this case an
oscillatory one.

\section{Graviatom existence conditions}

A graviatom can exist if the following conditions are fulfilled:

1) the geometrical condition $L>r_g+R$, where $L$ is the
characteristic size of the graviatom, $R$ is that of a charged
particle;

2) the stability condition: (a) $\tau_{\rm gr}<\tau_H$, where
$\tau_{\rm gr}$ is the graviatom lifetime, $\tau_H$ is the
mini-hole lifetime, (b) $\tau_{\rm gr}<\tau_p$, where $\tau_p$ is
the particle lifetime (for unstable particles);

3) the indestructibility condition (due to tidal forces and Hawking's
effect) $E_d < E_b$, where $E_d$ is the destructive energy, $E_b$ is the
binding energy.

If we introduce the dimensionless quantity
$$
    \alpha=\frac{GMm_p}{e^2},       \eqno(7)
$$
then the hydrogen-like case will correspond to
$$
    \alpha\ll \left(\frac{\hbar c}{e^2}\right)^2\frac{m_p}{mZ^2}  \eqno(8)
$$
 and the oscillatory one to
$$
    \alpha\gg \left(\frac{\hbar c}{e^2}\right)^2\frac{m_p}{mZ^2}.\eqno(9)
$$
The characteristic size of a hydrogen-like graviatom\\ $L=a_B^g$,
where
$$
    a_B^g=\frac{\hbar^2}{GMm^2}         \eqno(10)
$$
is Bohr's radius for the graviatom. The characteristic size of an
oscillatory graviatom is $L=r_Q$. The characteristic size of a nucleus is
$R=1.25\cdot 10^{-13}A^{1/3}$ cm, where A is the atomic weight of the
nucleus.

The graviatom lifetime is
$$
   \tau_{\rm gr}=\left(\frac{\hbar c}{e^2}\right)^5\left(\frac{m_p}{\alpha
        m}\right)^4\frac{\hbar}{Z^2mc^2}, \eqno(11)
$$
the  mini-hole lifetime is
$$
    \tau_H = \frac{15360\pi G^2M^3}{\hbar c^4}. \eqno(12)
$$
The lifetime of an unstable particle is
$$
    \tau_p = \frac{\hbar}{\Gamma},  \eqno(13)
$$
 where $\Gamma$ is the natural linewidth.

The destructive energy is $E_d = \{U_t,E_H\}$, with the tidal
energy
$$
    U_t=\frac{GMmR}{r^2} \eqno(14)
$$
and Hawking's radiation energy
$$
    E_H=\frac{b}{8\pi}\frac{\hbar c}{e^2}\frac{m_pc^2}{\alpha},\eqno(15)
$$
where $\hbar\omega_m = bkT,$ according to Wien's displacement law, and
$b=2.822$.

The binding energy is $E_b = \{E_W,I_{\rm ion}\}$, where $E_W$ is
the nuclear binding energy. The graviatom ionization energy is
$$
    I_{\rm ion}=\frac{m^3e^4\alpha^2}{2n^2\hbar^2m_p^2}. \eqno(16)
$$

In terms of $\alpha$, the graviatom existence conditions read
$$
    \alpha<\frac{1}{\sqrt{2}}\frac{\hbar c}{e^2}\frac{m_p}{m}
     \quad \ (a_B^g > r_g,\quad R\ll r_g), \eqno(17)
$$
$$
    \alpha>\left(\frac{\hbar
    c}{e^2}\right)^{8/7}\sqrt[7]{\frac{m_p^7}{15360\pi Z^2m^2_{pl}m^5}}\quad(\tau_{gr}<\tau_H), \eqno(18)
$$
 where $m_{\rm pl}=\sqrt{\hbar c/G}$ is the Planck mass,
$$
    \alpha>\left(\frac{\hbar
    c}{e^2}\right)^{5/4}\frac{m_p}{m}\sqrt[4]{\frac{\Gamma}{mc^2}}
        \frac{1}{\sqrt{Z}}\quad (\tau_{gr}<\tau_p), \eqno(19)
$$
$$
    \alpha>\frac{b}{8\pi}\frac{\hbar c}{e^2}\frac{m_pc^2}{E_W}
            \quad\quad (E_H<E_W), \eqno(20)
$$
$$
    \alpha>\frac{\hbar c}{e^2}\frac{m_p}{m}\sqrt[3]{\frac{bn^2}{4\pi}}
               \quad\quad  (E_H < I_{\rm ion}), \eqno(21)
$$
$$
    \alpha>\left(\frac{\hbar
    c}{e^2}\right)^{4/3}\frac{m_p}{m}\sqrt[3]
        {\frac{E_We^2}{m^2c^4R}}\quad (U_t<E_W),  \eqno(22)
$$
$$
    \alpha<\left(\frac{\hbar
    c}{e^2}\right)^2\frac{m_p}{m}\frac{e^2}{2n^2mc^2R}
    \quad (U_t<I_{\rm ion}).                      \eqno(23)
$$
\begin{figure}
\centering

\epsfig{file=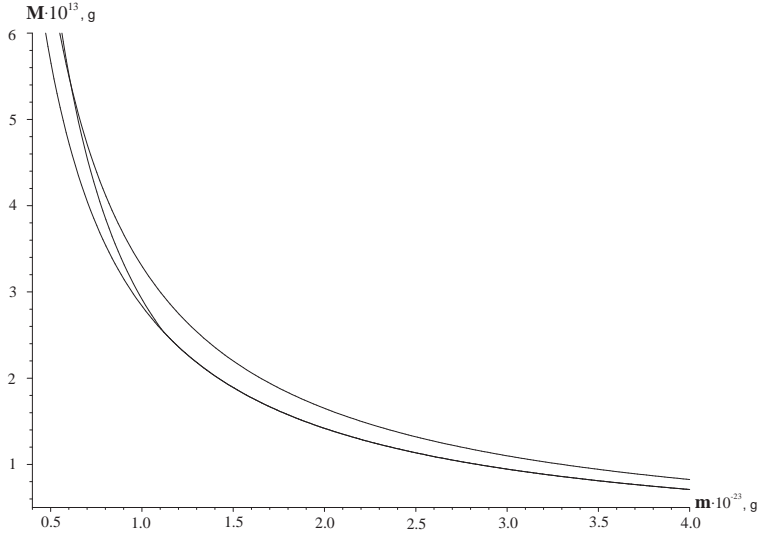,width=100mm}

\caption{\protect\small The dependence of mini-hole masses on the
charged particle masses satisfying the graviatom existence
conditions. The light curves indicate the range of values related
to the geometrical condition (the upper curve) and to Hawking's
effect ionization one (the lower curve). The heavy curve is
related to the particle stability condition ($\tau_p=10^{-22}$s).}
\medskip\hrule
\end{figure}

The conditions (17)--(23) prove to be fulfilled only for hydrogen-like
graviatoms, i.e., if the condition (8) is satisfied.  Moreover, it is the
graviatoms containing leptons and mesons with $Z=1$ that satisfy these
conditions. On the contrary, neither the graviatoms containing hadrons nor
atomic nuclei can satisfy them for both a hydrogen-like case (hadrons, light
nuclei) and an oscillatory one (heavy nuclei). The charged particles able to
be constituents of the graviatom are: the electron, muon, tau lepton, wino,
pion and kaon.

The graviatom existence conditions restrict the quantity $\alpha$
(proportional to the mini-hole mass $M$) within a narrow corridor
of its values depending on the charged particle mass $m$ (see
Fig.1).

\section{Graviatom radiation}

The intensity of the electric dipole radiation of a particle with
mass $m$ and charge $e$ in the gravitational field of a mini-hole
reads
$$
    I^d_{fi} = \frac{2\hbar e^2\omega_{if}^3f_{if}}{mc^3}, \eqno(24)
$$
where $\omega_{if} = (E_i-E_f)/\hbar$ is the frequency of the
transition $i\to f$ and $f_{if}$ is the oscillator strength
\cite{BS}. The condition of radiation being dipole is
$$
    \alpha < \frac{m_p}{m}\frac{\hbar c}{e^2}, \eqno(25)
$$
which almost coincides with the geometrical condition (17) to be
satisfied for the graviatoms under consideration.


\begin{table*}
\begin{center}
Table 1. Graviatom parameters for the electron, muon and tau
lepton.

\begin{tabular}{ |l|l|l|l|l|l|l| }
\hline & \multicolumn{6}{|c|}{Charged particles} \\
\cline{2-7} & \multicolumn{2}{|c|}{ e } & \multicolumn{2}{|c|}{ $\mu$ } & \multicolumn{2}{|c|}{ $\tau$ }\\
\hline
$mc^2$,  {MeV} & \multicolumn{2}{|l|}{ 0.511 } & \multicolumn{2}{|l|}{ 105.659 } & \multicolumn{2}{|l|}{ 1.777E+3 }\\
$\tau_p$,  {s} & \multicolumn{2}{|l|}{ $\infty$ } & \multicolumn{2}{|l|}{ 2.200E-6 } & \multicolumn{2}{|l|}{ 2.900E-13 }\\
\cline{2-7}
 & min & max & min & max & min & max\\
\cline{2-7}
M,  {g} & 3.12E+17 & 3.62E+17 & 1.51E+15 & 1.75E+15 & 9.96E+13 & 1.04E+14\\
$r_g$,  {cm} & 4.69E-11 & 5.46E-11 & 2.27E-13 & 2.64E-13 & 1.35E-14 & 1.57E-14 \\
$a_B^g$,  {cm} & 5.46E-11 & 6.35E-11 & 2.64E-13 & 3.07E-13 & 1.57E-14 & 1.83E-14 \\
$\hbar\omega_{12}$,  {MeV} & 0.071 & 0.096 & 14.64 & 19.81 & 246.2 & 333.2 \\
$\hbar\omega_{13}$,  {MeV} & 0.084 & 0.114 & 17.35 & 23.48 & 291.8 & 394.9 \\
$I^{d}(2p\to 1s)$,  {erg $s^{-1}$} & 1.02E+10 & 2.55E+10 & 4.39E+14 & 1.09E+15 & 1.24E+17 & 3.08E+17 \\
$I^{d}(3p\to 1s)$,  {erg $s^{-1}$} & 3.25E+9 & 8.05E+9 & 1.39E+14 & 3.44E+14 & 3.93E+16 & 9.74E+16 \\
$I^{q}(3d\to 1s)$,  {erg $s^{-1}$} & 1.33E+8 & 4.47E+8 & 5.70E+12 & 1.91E+13 & 1.61E+15 & 5.41E+15 \\
$I^{g}(3d\to 1s)$,  {erg $s^{-1}$} & 1.11E+10 & 4.36E+10 & 4.75E+14 & 1.85E+15 & 1.34E+17 & 5.24E+17 \\
$E_H$,  {MeV} & 0.081 & 0.094 & 16.78 & 19.52 & 282.2 & 328.3 \\
$P_H$,  {erg $s^{-1}$} & 2.63E+10 & 3.56E+10 & 1.13E+15 & 1.52E+15 & 3.18E+17 & 4.31E+17 \\
\hline
\end{tabular}
\end{center}
\end{table*}

\begin{table*}
\begin{center}
Table 2. Graviatom parameters for the wino, pion and kaon.

\begin{tabular}{ |l|l|l|l|l|l|l| }
\hline  & \multicolumn{6}{|c|}{Charged particles} \\
\cline{2-7}  & \multicolumn{2}{|c|}{ $\tilde W$ } & \multicolumn{2}{|c|}{ $\pi$ } & \multicolumn{2}{|c|}{ K }\\
\hline
$mc^2$,  {MeV} & \multicolumn{2}{|l|}{ 8.000E+5 } & \multicolumn{2}{|l|}{ 139.568 } & \multicolumn{2}{|l|}{ 493.994 }\\
$\tau_p$,  {s} & \multicolumn{2}{|l|}{ 5.000E-10 } & \multicolumn{2}{|l|}{ 2.600E-8 } & \multicolumn{2}{|l|}{ 1.200E-8 }\\
\cline{2-7}
 & min & max & min & max & min & max \\
\cline{2-7}
M,  {g} & 1.99E+11 & 2.31E+11 & 1.14E+15 & 1.33E+15 & 3.22E+14 & 3.75E+14 \\
$r_g$,  {cm} & 2.99E-17 & 3.49E-17 & 1.72E-13 & 1.99E-13 & 4.86E-14 & 5.65E-14 \\
$a_B^g$,  {cm} & 3.49E-17 & 4.06E-17 & 1.99E-13 & 2.33E-13 & 5.65E-14 & 6.57E-14 \\
$\hbar\omega_{12}$,  {MeV} & 1.11E+5 & 1.50E+5 & 19.34 & 26.17 & 68.44 & 92.62 \\
$\hbar\omega_{13}$,  {MeV} & 1.31E+5 & 1.78E+5 & 22.92 & 31.02 & 81.12 & 109.78 \\
$I^{d}(2p\to 1s)$,  {erg $s^{-1}$} & 2.52E+22 & 6.24E+22 & 7.66E+14 & 1.90E+15 & 9.60E+15 & 2.38E+16 \\
$I^{d}(3p\to 1s)$,  {erg $s^{-1}$} & 7.96E+21 & 1.97E+22 & 2.42E+14 & 6.01E+14 & 3.04E+15 & 7.53E+15 \\
$I^{q}(3d\to 1s)$,  {erg $s^{-1}$} & 3.27E+20 & 1.10E+21 & 9.95E+12 & 3.34E+13 & 1.25E+14 & 4.18E+14 \\
$I^{g}(3d\to 1s)$,  {erg $s^{-1}$} & 2.72E+22 & 1.06E+23 & 8.29E+14 & 3.23E+15 & 1.04E+16 & 4.05E+16 \\
$E_H$,  {MeV} & 1.27E+5 & 1.48E+5 & 22.16 & 25.78 & 78.44 & 91.25 \\
$P_H$,  {erg\ s$^{-1}$} & 6.45E+22 & 8.73E+22 & 1.96E+15 & 2.66E+15 & 2.46E+16 & 3.33E+16 \\
\hline
\end{tabular}
\end{center}
\end{table*}

\begin{table*}
\begin{center}
Table 3. Relations valid for all graviatoms

\begin{tabular}{ |l|l|l| }
\hline
 & \multicolumn{2}{|c|}{ Values } \\
\cline{2-3}
\multicolumn{1}{|c|}{ Relations } & \multicolumn{1}{|c|}{ min } & \multicolumn{1}{|c|}{ max } \\
\hline
\qquad $I^{g}(3d\to 1s) / I^q(3d\to 1s)$ \qquad & \quad 83.295 \qquad & \quad 96.899 \qquad \\
\qquad $I^{g}(3d\to 1s) / I^d(2p\to 1s)$ \qquad & \quad 1.082 & \quad 1.703 \qquad \\
\qquad $I^{g}(3d\to 1s) / I^d(3p\to 1s)$ \qquad & \quad 3.419 & \quad 5.383 \qquad \\
\qquad $I^d(2p\to 1s)  / P_H$ \qquad & \quad 0.390 \qquad & \quad 0.715 \qquad \\
\qquad $\hbar\omega_{12} / E_H$ \qquad & \quad 0.872 \qquad & \quad 1.015 \qquad \\
\qquad $\sqrt{Mm} / m_{pl}$ \qquad & \quad 0.780 \qquad & \quad 0.841 \qquad \\
\hline
\end{tabular}
\end{center}
\end{table*}


The electric dipole radiation intensity for the hydrogen-like
graviatom performing the transition $2p\to 1s$ is
$$
    I^d_{12}=\frac{2\hbar e^2\omega^3_{21}}{mc^3}f_{2p\to 1s}, \eqno(26)
$$
with the oscillator strength
$$
f_{2p\to 1s}=\frac{2^{13}}{3^9}=0.4162 \eqno(27)
$$
and the transition energy
$$
    \hbar\omega_{12} = \frac{3\alpha^2e^4m^3}{8m_p^2\hbar^2}. \eqno(28)
$$
Finally we obtain
$$
I^d_{12}=\frac{2^5\alpha^6e^{12}m^8}{3^6c^3\hbar^8m_p^6}.\eqno(29)
$$
The electric dipole radiation intensity for the transition $3p\to
1s$ is
$$
I^d_{13}=\frac{2\hbar e^2\omega^3_{31}}{mc^3}f_{3p\to 1s}
\eqno(30)
$$
with the oscillator strength
$$
f_{3p\to 1s}=\frac{3^4}{2^{10}}=0.0791 \eqno(31)
$$
and the transition energy
$$
\hbar\omega_{31}=\frac{2^2\alpha^2e^4m^3}{3^2\hbar^2 m_p^2}.
\eqno(32)
$$
Finally we obtain
$$
I^d_{13}=\frac{\alpha^6e^{14}m^8}{2^33^2c^3\hbar^8m_p^6}.
\eqno(33)
$$
Hence it follows
$$
\frac{I^d_{12}}{I^d_{13}}=\frac{2^8}{3^4}=3.161,
\quad\frac{\omega_{21}}{\omega_{31}}=\frac{3^3}{2^5}=0.844.
$$
The electric quadrupole radiation intensity for the transition
$3d\to 1s$ is
$$
I^q_{13}=\frac{6\hbar e^2\omega^3_{31}}{mc^3}f_{3d\to 1s}
\eqno(34)
$$
with the oscillator strength
$$
f_{3d\to 1s}=\frac{3^7}{2^{16}}\left(\frac{Mm}{m^2_{pl}}\right)^2,
\eqno(35)
$$
where the transition energy $\omega_{31}$ is given by formula
(32). Finally we obtain
$$
I^q_{13}=\frac{\alpha^8e^{18}m^{10}}{2^33^4c^5\hbar^{10}m_p^8}.\eqno(36)
$$
The gravitational radiation intensity for the graviatom performing
the transition $3d\to 1s$ reads
$$
I^g_{13}=\frac{6\hbar GM\omega^3_{31}}{c^3}f_{3d\to 1s}, \eqno(37)
$$
where the oscillator strength and the transition energy are given
by formulae (32) and (35) respectively. Finally we obtain
$$
I^g_{13}=\frac{\alpha^9e^{18}m^{11}}{2^33^4c^5\hbar^{10}m_p^9}.
\eqno(38)
$$

 The mini-hole creates particles near its horizon
which can ionize graviatoms and split nuclei being their
constituents. Hawking's effect power is given by the formula
$$
    P_H=\frac{1}{15360\pi \alpha^2}\left(\frac{\hbar
            c}{e^2}\right)^2\frac{(m_pc^2)^2}{\hbar}. \eqno(39)
$$
Hawking's energy is
$$
    E_H=\frac{b}{8\pi\alpha}\frac{\hbar c}{e^2}m_pc^2. \eqno(40)
$$

The mini-hole mass $M$ is diminishing due to Hawking's effect
evaporation, i.e.
$$
M_f=\sqrt[3]{M_i^3-\frac{\hbar c^4}{5120\pi HG^2}}, \eqno(41)
$$
where $M_f$ and $M_i$ are the final and initial mini-hole masses
respectively, $H$ is the Hubble  parameter. $M_f=0$ corresponds to
the minihole mass $M_f$ to have evaporated for the Universe's
lifetime $T$
$$
M_i=\sqrt[3]{\frac{\hbar c^4}{5120\pi HG^2}}, \eqno(42)
$$
which gives $M_i=4.38\cdot 10^{14}g$ for $H=65 kms^{-1}Mpc^{-1}
\quad(H=\frac{1}{T})$.

 Tables 1 and 2 present the graviatom
parameters: the mini-hole and charge particle masses satisfying
the graviatom existence conditions, the energies and intensities
of the electromagnetic, gravitational and Hawking's radiations.
Besides, unstable particle lifetimes (for the wino see \cite{S})
and Bohr's graviatom radii are indicated.

The mini-holes being constituents of the graviatoms are formed due
to Jeans' gravitational instability at the times about
$\frac{r_g}{c}=10^{-27}\div 10^{-21}s$ from the initial
singularity. The mini-hole masses for the graviatoms involving
electrons, muons and pions exceed the value of $4.38\cdot 10^{14}
g$, which means that it is possible for such graviatoms to have
existed up to now \cite{F}.

 Table 3 presents relations valid for all graviatoms: the
gravitational-to-electromagnetic radiation intensity ratios, the
dipole-to-Hawking radiation ratio as well as the quantity equal to
the square root of $\frac{GMm}{\hbar c}=0.608\div 0.707 $. The
latter is a gravitational equivalent of the fine structure
constant. The gravitational radiation intensities two orders
exceed the electromagnetic ones. The graviatom dipole radiation
energies and intensities have proved to be comparable with those
for Hawking's effect of the mini-holes being constituents of the
graviatoms. The gravitational equivalent of the fine structure
constant does not exceed unity, thus the perturbation theory
remains valid.

\section{Conclusion}

We have considered the graviatom existence conditions proving to
be satisfied for charged leptons and mesons but not baryons
(protons and nuclei). The baryon sizes appear to exceed the
mini-hole gravitational radii which means that neither a
hydrogen-like nor oscillatory case can take place, i.e., stable
graviatoms with baryon constituents become impossible. Instead of
them, there occurs a so-called quantum accretion of baryons onto a
mini-hole. The internal structure of the baryons, consisting of
quarks and gluons, should be taken into account. The whole problem
is solvable within the framework of quantum chromodynamics and
quantum electrohydrodynamics. The radiation of baryons and quarks
is also worth consideration later on. In the future, also of
interest is to consider the graviatoms as sources of the
electromagnetic background radiation and their possible
contribution to dark matter.

\end{document}